\documentclass{aip-cp}

\usepackage[numbers]{natbib}
\usepackage{rotating}
\usepackage{graphicx}
\newcommand*\savesymbol[1]{%
  \expandafter\let\csname orig#1\expandafter\endcsname\csname#1\endcsname
  \expandafter\let\csname #1\endcsname\relax
}

\newcommand*\restoresymbol[2]{%
  \expandafter\global\expandafter\let\csname#1#2\expandafter\endcsname%
    \csname#2\endcsname
  \expandafter\global\expandafter\let\csname#2\expandafter\endcsname%
    \csname orig#2\endcsname
}

\savesymbol{iint}
\savesymbol{iiint}
\savesymbol{iiiint}
\savesymbol{idotsint}
\usepackage{amsmath}
\usepackage{amssymb}
\usepackage{bm}
\usepackage{xspace}

\newcommand{\DDstar}{\ensuremath{\bar{D}^0 D^{*0}}\xspace}
\newcommand{\kmax}{\ensuremath{k_0^\text{max}}\xspace}
\newcommand{\lhcb}{LHCb\xspace}
\newcommand{\cdf}{CDF\xspace}
\newcommand{\cms}{CMS\xspace}
\newcommand{\Dzero}{D$\varnothing$\xspace}
\newcommand{\alice}{ALICE\xspace}
\newcommand{\helium}{\ensuremath{^3\text{He}}\xspace}
\newcommand{\hypertriton}{\ensuremath{{}_\Lambda^3\text{H}}\xspace}
\newcommand{\jpsi}{\ensuremath{{J\mskip -3mu/\mskip -2mu\psi\mskip 2mu}}\xspace}
\newcommand{\ie}{{\em i.e.}\xspace}
\newcommand{\tev}{\ensuremath{\text{TeV}}\xspace}

\begin{document}

% Title portion
\title{Production of exotic hadrons at hadron colliders}

 \author[aff1,aff2]{Alessandro Pilloni\corref{cor1}}
 
 \affil[aff1]{``Sapienza'' Universit\`a di Roma, Dipartimento di Fisica and INFN\\
 P.le Aldo Moro 5, I-00185 Roma (Italy)}
 \affil[aff2]{Theory Center, Thomas Jefferson National Accelerator Facility,\\
 12000 Jefferson Avenue, Newport News, VA 23606, USA}
 \corresp[cor1]{Corresponding author: alessandro.pilloni@roma1.infn.it}

\maketitle

\begin{abstract}
The observation of many unexpected states decaying into heavy quarkonia has challenged the usual $Q \bar Q$ interpretation. 
One of the most studied exotic states, the $X(3872)$, happens to be copiously produced in high-energy hadron collisions. We discuss how this large prompt production cross-section disfavors a loosely-bound molecule interpretation for this particle. This is supported by Monte Carlo simulations, and by a comparison with extrapolated light nuclei data by \alice.
\end{abstract}

\section{Introduction}
In the last ten years lots of unexpected $XYZ$ resonances have been discovered in the heavy quarkonium sector.  
Their production and decay rates are not compatible with a standard quarkonium interpretation. For an extended review on this topic, see~\cite{Esposito:2014rxa,Faccini:2012pj}. The most popular phenomenological interpretation for many of these states is the so-called {\em hadron molecule}, \ie a loosely bound state of two mesons, interacting via long-range light meson exchange.
The main antagonistic model is the {\em compact tetraquark}, which is dominated by short-range color interaction~\cite{Maiani:2004vq,Faccini:2013lda,Maiani:2014aja}. 

\section{Prompt production of $X(3872)$ at hadron colliders}
\label{sec:pioni}
The $X(3872)$ is known to have a large prompt~\footnote{\ie not produced at the displaced $B$ or $\Lambda_b$ decay vertex, but at the hadron collision vertex.}  production cross section, both in $pp$~\cite{Aaij:2011sn,Chatrchyan:2013cld} and $p\bar p$~\cite{Abazov:2004kp,cdfnote} collisions, of the same order of magnitude of the ordinary $\psi(2S)$ charmonium state.

The closeness of the $X(3872)$ to the $DD^*$ threshold suggests for this state to be a \DDstar molecule~\footnote{The charge-conjugate mode is understood}, with a binding mechanism provided by some kind of inter-hadron potential,  e.g. one-pion exchange. This interpretation would require a negative binding energy, which cannot be confirmed with the current experimental precision (\mbox{$\mathcal{E}_b = M_{X(3872)} - M_{D^0} - M_{D^{*0}} \simeq -3 \pm 192$~keV}~\cite{Tomaradze:2015cza}). Moreover, if the molecular hypothesis holds, the partial width of  $X(3872)\to \DDstar$ can be calculated as a function of the binding energy~\cite{Polosa:2015tra}, and future precise measurements of the mass, width and branching ratios of the $X(3872)$ will therefore allow to test this model.

Anyway, it is hard to explain how a loosely bound molecule, with binding energy compatible with zero, could be formed within the hadrons ejected in hadron collisions at energies of some TeV. This issue was firstly raised in~\cite{Bignamini:2009sk}. Firstly, it was considered that, if the $X(3872)$ is a molecule, and its wave function is negligible if the relative momentum $k_0$ is larger than some \kmax, then the number of $X$ is bounded by the number of \DDstar pairs with $k_0 < \kmax$. Using standard hadronization algorithms (Herwig and Pythia), 
 the production of open charm mesons was simulated, and the number of \DDstar pairs was studied as a function of their relative $k_0$. If one considers a \mbox{$\kmax \sim 50$~MeV}, the upper bound for the  prompt production cross section of the $X(3872)$ is found to be $\sim 300$~times smaller than the experimental value, thus challenging the molecular interpretation of such a state. 

The authors of~\cite{Artoisenet:2009wk} proposed that the final state interactions between the two mesons, in the Migdal-Watson framework, might increase the cross section. Indeed, they evaluated an enhancement factor, which gives a factor of few only, but they argue that the very presence of strong rescattering allows for a  $\kmax \sim O(m_\pi)  \lesssim 500$~MeV without breaking the bound state. Since the cross section scales as $(\kmax)^3$, the experimental value is rapidly exceeded. This approach was criticised in~\cite{Bignamini:2009fn}, by noting that the presence of a huge numbers of pions between the $DD^*$ mesons make the Migdal-Watson approach unjustified. The controversy remained somewhat unsolved~\cite{Artoisenet:2010uu}. The same Migdal-Watson approach has been recently invoked to estimate the prompt production cross section of $Z_{c,b}$~\cite{Guo:2013ufa}, $X_b$~\cite{Guo:2014sca}, $D_{sJ}$~\cite{Guo:2014ppa} exotic states. However, the application of this formalism to the above-threshold charged states is unclear, and the connection of \kmax with the natural width of the state looks quite arbitrary; in the absence of precise final state interactions calculations, the uncertainty in \kmax reflects in a $> O(10)$ uncertainty in the cross sections, which makes any estimate of sizeable cross sections unreliable~\cite{Guo:2013ufa,Guo:2014sca,Guo:2014ppa}.

\begin{figure}[t]
\centering
\includegraphics{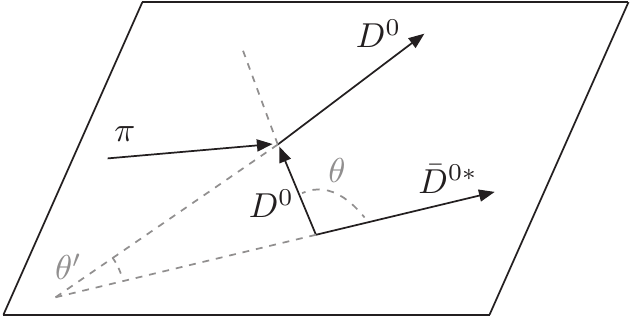}
\caption{The elastic scattering of a $\bar{D}^0$ (or $D^{*0}$) with a pion among those produced in hadronization could reduce the relative momentum $k_0$ in the centre of mass of the \DDstar pair~\cite{Esposito:2013ada,Guerrieri:2014gfa}.  }
\label{fig:angoli}
\end{figure}

In Ref.~\cite{Esposito:2013ada} we suggested a more mechanistic way to take into account final state interactions: it was considered that some of the large number of pions produced in the neighbourhood of the open charm meson pairs could scatter  elastically on the $\bar{D}^0$ or $D^{*0}$ component of the would-be-molecule, thus changing the relative momentum in the centre of mass of the pair, $k_0$  (Fig.~\ref{fig:angoli}).
If this interaction reduces the relative momentum of even a small part of the many large-$k_0$ pairs, there could be a significant effect of feed-down of pairs towards lower bins, even in the far low energy region below 50~MeV. Populating that region means increasing the formation probability of the loosely bound $X$. 
A first qualitative exploration of this phenomenon was performed, by generating samples of  $p\bar p\to  c\bar c \,(X)$ events in Herwig and Pythia, at Tevatron COM energies $\sqrt{s}=1.96$~TeV. Then, a pion in proximity of one of the mesons is selected, the elastic interaction (regulated by effective matrix elements) is implemented. It was noticed that the number of pairs with $k_0 < k_0^\text{max}$ was sensibly increased. The interactions was repeated three times using a Markov chain-like formalism. The result was striking, with the $k_0 < 50$~MeV bin increasing by two orders of magnitude.

\begin{figure}[t]
\centering
 \includegraphics[width=.48\textwidth]{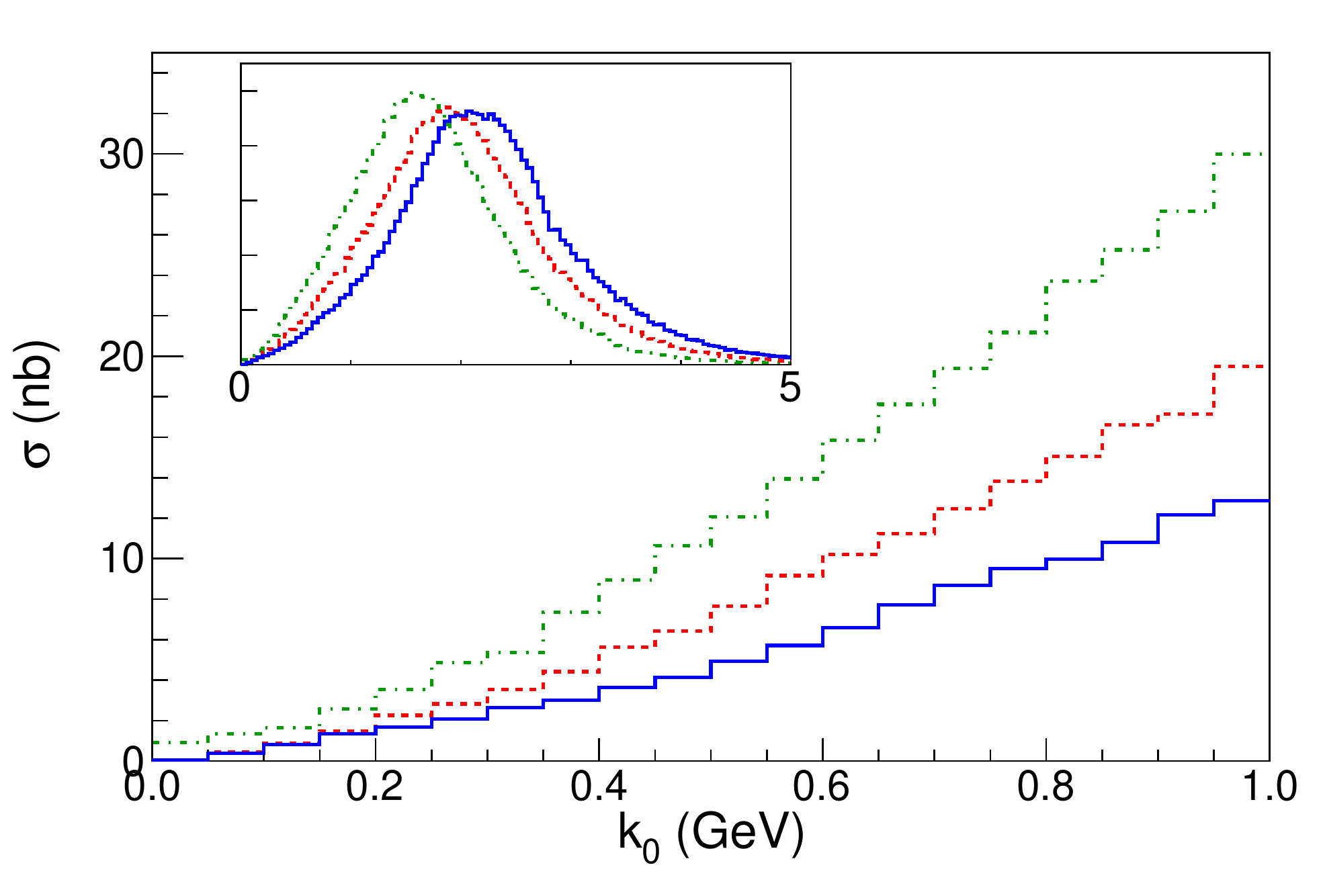}
\caption{Integrated cross sections of \DDstar, without (blue, solid), with one (red, dashed) and three (green, dot-dashed) interacions with pions, with the method of~\cite{Guerrieri:2014gfa}. In the inset the same plot on a wider range of $k_0$ values. }
\label{markov3}
\end{figure}

The complete analysis of the full QCD events $p \bar p\to c\bar c, p \bar p\to gg,p \bar p\to g q,p \bar p\to q q ...$ was presented in Ref.~\cite{Guerrieri:2014gfa}.
We generated $9\times 10^{10}$ events in Herwig and $5\times 10^{10}$ events in Pythia. First of all, we checked that the interaction with pions does not spoil the known $D$ meson distributions. However, we noticed that in the full QCD simulations more than $95\%$ of $c \bar c$ pairs are produced during the parton shower process by softer and softer gluons, so that the number of low-relative-momentum $DD^*$ pairs is larger with respect to $p \bar p \to c \bar c$ events.
We show the results in Fig.~\ref{markov3} and Tab.~\ref{tablemarkov}. 
The enhancement due to interactions with pions is still present, but is not as dramatic as in $c \bar c$ generation. In fact, we increase the cross section only by a factor $\mathcal{O}(10)$, instead that $\mathcal{O}(100)$ as in~\cite{Esposito:2013ada}. Even if we take into account a $k_0^\text{max} = 100$~MeV, and if we consider $3\div 5$ interactions with pions, we can get only up to $30 \%$ of experimental cross section. 

\begin{table}[b]
\begin{tabular}{lccc}\hline
$k_0^\text{max}$ & 50 MeV & 300 MeV & 450 MeV \\ \hline
$\sigma(0\pi)$ & 0.06 nb & 6 nb & 16 nb \\
$\sigma(1\pi)$ & 0.06 nb & 8 nb & 22 nb \\
$\sigma(3\pi)$ & 0.9 nb & 15 nb & 37 nb \\ \hline
\end{tabular}
\caption{Effect of multiple scattering in $X(3872)$ production cross section (see Fig.~\ref{markov3})~\cite{Guerrieri:2014gfa}. $k_0^{\rm max}$ indicates the integration range $[0,k_0^{\rm max}]$.}\label{tablemarkov}
\end{table}

\section{Comparison of $X(3872)$ with light nuclei production}
Very recently the \alice collaboration reported results on the production of deuteron, helium-3 (\helium) 
and hypertriton (\hypertriton) light nuclei in relatively high $p_\perp$ bins in Pb-Pb 
collisions, at $\sqrt{s_\text{NN}}=2.76$~TeV~\cite{Adam:2015yta,Adam:2015vda}. If the $X(3872)$ shares the same molecule nature with light nuclei, one expects similar production cross sections, especially at high values of $p_\perp$.

\begin{figure}[t]
\centering
  \includegraphics[width=.48\textwidth]{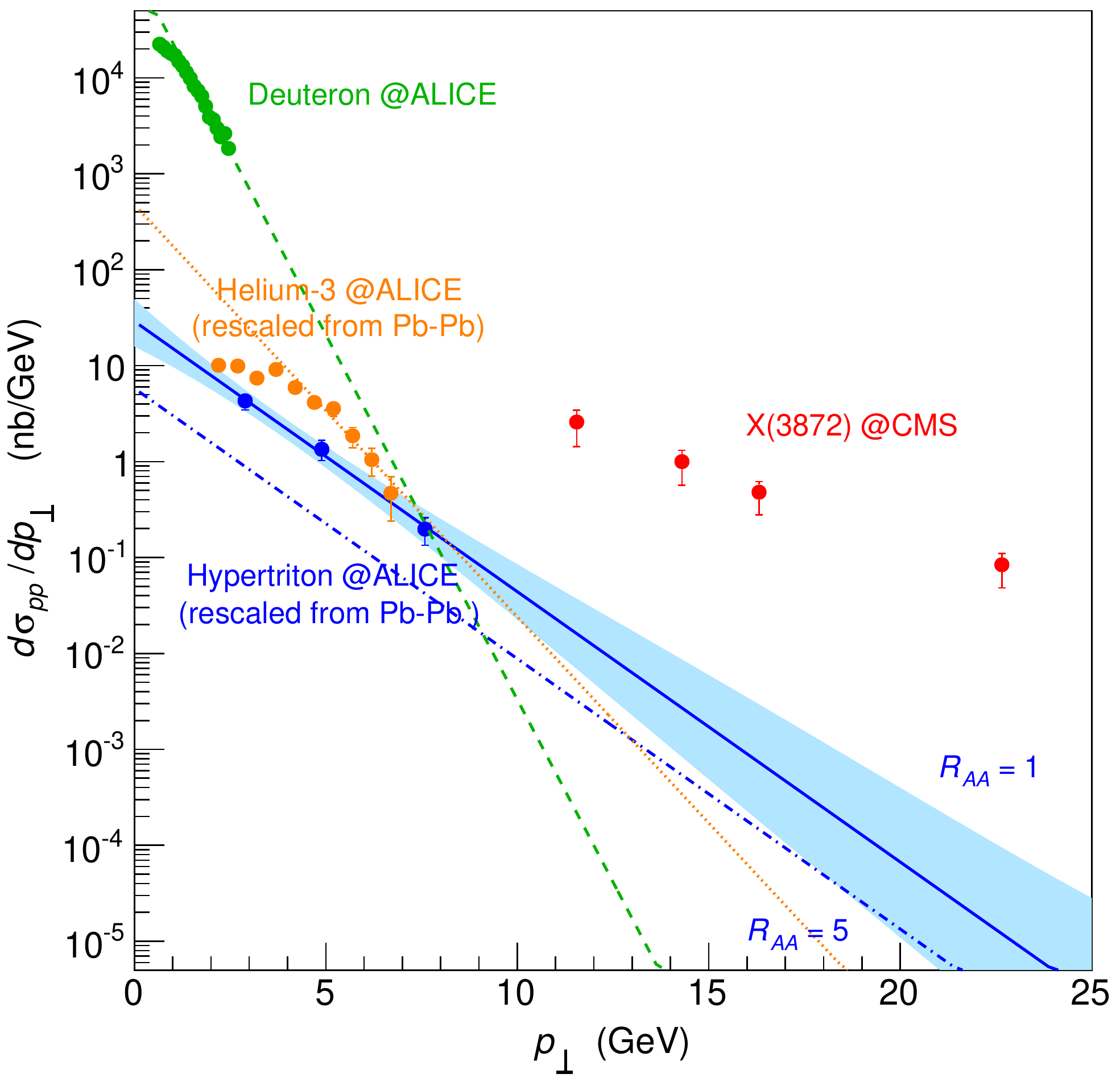} 
  \includegraphics[width=.48\textwidth]{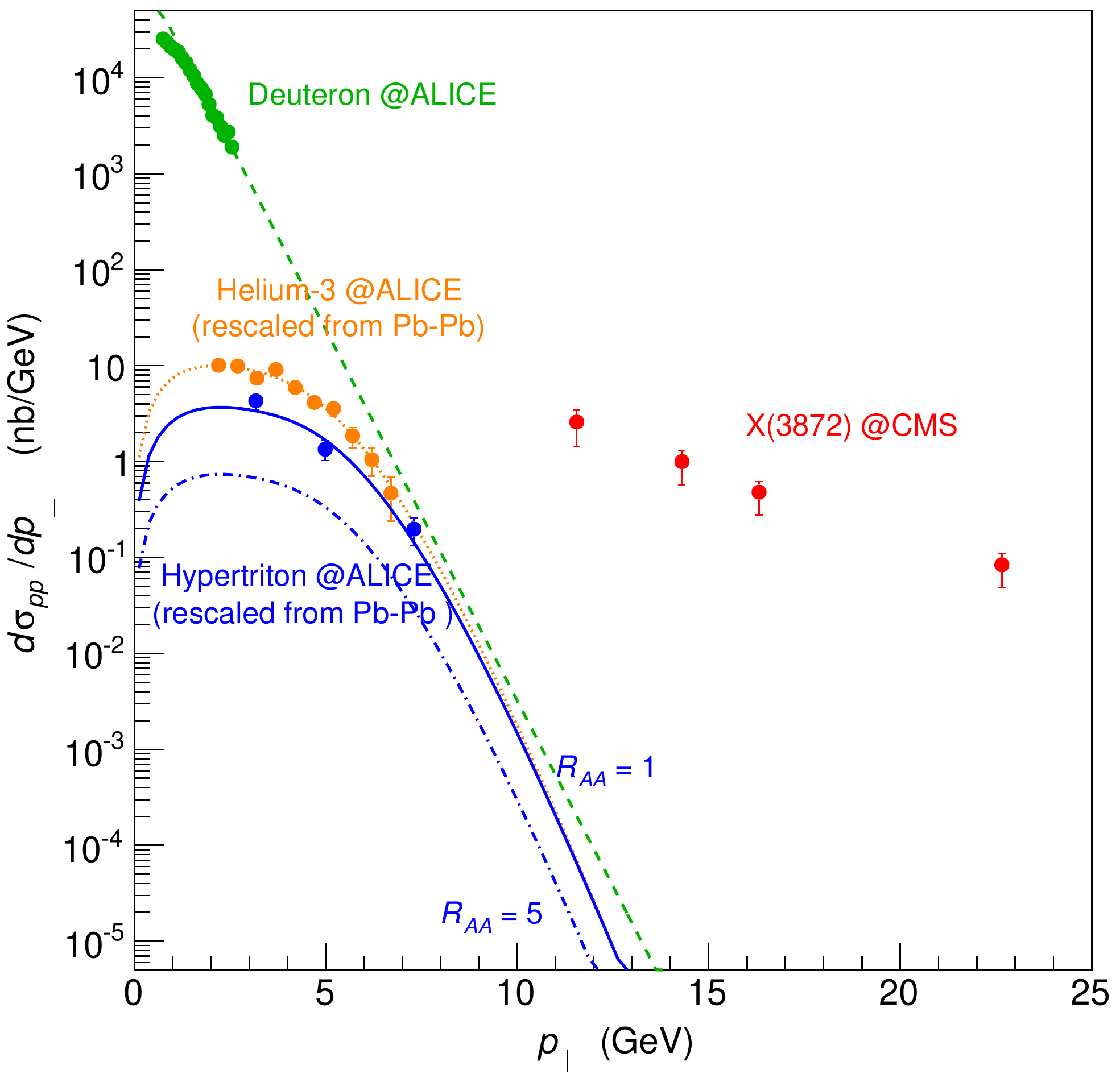}
\caption{Comparison between the prompt production cross section in $pp$ collisions of $X(3872)$ (red), 
deuteron (green), \helium (orange), and hypertriton (blue)~\cite{Esposito:2015fsa}. The $X$ data from \cms~\cite{Chatrchyan:2013cld} are rescaled 
by the branching ratio $\mathcal{B}(X\to \jpsi\,\pi\pi)$. Deuteron data in $pp$ collisions are taken from 
\alice~\cite{Adam:2015vda}.
The \helium and hypertriton data measured by \alice in Pb-Pb collisions~\cite{Adam:2015yta,Adam:2015vda} have been 
rescaled to $pp$ using a Glauber model, as explained in the text. 
The dashed green line is the exponential fit to the deuteron data points in the $p_\perp\in[1.7,3.0]$ GeV region, 
whereas the dotted orange one is the fit to the \helium data points.  The solid and dot-dashed blue lines represent 
the fits to hypertriton data with $R_{AA}=1$ (no medium effects) and an hypothetical constant value of $R_{AA}=5$. \label{uno}
({Left Panel}) The hypertriton data are fitted with an exponential curve, and the light blue band is the 
$68\%$~C.L. for the extrapolated $R_{AA}=1$ curve. \helium data in the $p_\perp \in[4.45,6.95]$ GeV region are 
also fitted with an exponential curve. 
({Right Panel}) The hypertriton and \helium data are fitted with blast-wave functions.}
\end{figure}

Although a proper comparison would require the measurement of light nuclei production in $pp$ collisions only, and at the same $p_\perp > 15$~GeV where the $X$ is seen, simple extrapolation of available data can been performed to get some qualitative information~\cite{Esposito:2015fsa}.
As a first approximation one can assume that there are no medium effects enhancing or suppressing the 
production of light nuclei in Pb-Pb collisions. This is equivalent to state that each nucleus-nucleus collision is 
just an independent product of $N_\text{coll}$ proton-proton collisions, where $N_\text{coll}$ is computed in a Glauber 
Monte Carlo calculation as a function of the centrality class. For example, the hypertriton production cross section in $pp$ collisions 
can be estimated to be 
\begin{equation}
\label{eq:glauber}
\left(\frac{d\sigma\left(\hypertriton\right)}{dp_\perp}\right)_{pp}=\quad\frac{\Delta y}{{\cal B} (\helium\,\pi)} 
\times
\frac{\sigma_{pp}^\text{inel}}{N_\text{coll}^{0-10\%}}\left(\frac{1}{N_{\text{evt}}}\frac{d^2N(\helium\,\pi)}{dp_\perp dy}
\right)_\text{Pb-Pb}^{0-10\%},
\end{equation}
where the use of the factor $\sigma_{pp}^\text{inel}(\sqrt{s}=7~\tev)$ understands a naive rescaling from $\sqrt{s}= 2.76$~TeV to $\sqrt{s} = 7$~TeV.
Similarly, we can estimate the \helium distribution in $pp$ collisions from the \alice Pb-Pb data in 
the $0$-$20\%$ centrality class.
The deuteron has been measured directly in $pp$ collisions at $\sqrt{s}=7$~TeV, so no calculation is needed.
To extrapolate the data points towards higher values of $p_\perp$, we perform exponential fits. Alternatively, we fit hypertriton and \helium data with the blast-wave 
model~\cite{Schnedermann:1993ws}, which is expected to reproduce correctly the low and medium $p_\perp$ 
regions in Pb-Pb collisions. Since we are rescaling Pb-Pb data to $pp$ by a constant factor, the same shape 
holds in our estimated $pp$ data, and gives a guess on the asymptotic exponential behavior. 
The results are shown in Fig.~\ref{uno}.

The \cms analysis of $X$ production provides the differential cross section times the branching 
fraction $\mathcal{B}\left(X(3872) \to \jpsi\,\pi^+\pi^-\right)$, for which we used the value $\mathcal{B} = 8.1^{+1.9}_{-3.1}\%$~\cite{Esposito:2014rxa}.  
The comparison in Fig.~\ref{uno} shows that the extrapolated hypertriton production cross section in $pp$ collisions would 
fall short by about $2\div 3$ orders of magnitude with respect to the $X$ production, and much more 
according to the blast-wave fit in the right panel. The drop of the deuteron cross section, 
which is directly measured in $pp$ collisions, appears definitely faster. 

As we discussed, the main problem for the production of loosely bound molecular states in 
proton-proton collisions is the difficulty in  producing the constituents close enough in phase space. 
However, it is well known that the interaction of elementary partons with the collective hot dense 
medium causes relevant energy loss of the partons themselves. This effect is usually quantified by 
the nuclear modification factor $R_{AA}$,
which compares the particle yield in Pb-Pb collisions with that in $pp$. It then follows that the method used 
to obtain Eq.~(\ref{eq:glauber}) corresponds  to assume $R_{AA}=1$.

While for ordinary hadrons medium effects generally lead to a suppression of the particle yield, \ie 
$R_{AA}<1$, conversely molecular states with small binding energy are expected to 
be \emph{enhanced}, \ie $R_{AA}>1$~\cite{Cho:2010db}. The role of the medium would be, 
in fact, that of decreasing the relative momenta of the components with respect to the zero temperature case 
due to the well-known jet quenching effect~\cite{Gyulassy:1990ye}. This would favor their coalescence 
into the final bound state by reducing their relative momenta directly at parton level. 
Unfortunately there is no measurement of $R_{AA}$ for the deuteron as a function of $p_\perp$, but a naive estimate based on available \alice data suggests for values $R_{AA}\sim 5$ at $p_\perp = 5$~GeV.
Moreover, one can use \alice deuteron data to calculate $R_{CP}$, \ie the ratio between central and peripheral collisions, which is strongly correlated to $R_{AA}$.
We have $R_{CP} = 1.7$ at the last point with $p_\perp = 3.1$~GeV, and $R_{CP}$ keeps increasing at larger $p_\perp$ according to the blast-wave fitting function.
This confirms the enhancement for the production of hadron molecules.
One naturally expects such an enhancement to be even more relevant for $3$-body nuclei like \helium 
and the hypertriton. Its role would be to further \emph{decrease} the extrapolated cross section in prompt 
$pp$ collisions. As we already said, indeed, a value of $R_{AA} > 1$ applied to Pb-Pb data implies a $pp$ cross section even smaller than predicted by the Glauber model. 
Even though qualitative conclusions can already be drawn, a quantitative analysis 
substantiated by data at higher $p_\perp$ is necessary for a definitive comparison with the $X$ case. 

\section{The diquark-antidiquark description and Hadronization}
Because of previous analyses, we are led to conclude that the hadronization of multiquark hadrons in prompt collisions at LHC must proceed through the formation of compact clusters of quarks,
with color neutralized in all possible ways.
The two-meson states $|(Q\bar q)_{\bm 1_c}(\bar Q q)_{\bm 1_c}\rangle$ and $|(Q\bar Q)_{\bm 1_c}(q\bar  q)_{\bm 1_c}\rangle$ will tend to fly apart, because the residual strong force is not sufficient to produce bound states.  In this sense such  states are in  the continuum spectrum of an `open channel'  potential. A diquark-antidiquark state is instead kept together by color interactions, which are very likely to produce a discrete spectrum (closed channel). The generic state can be therefore considered as a superposition
\begin{equation}
|\psi\rangle =\alpha |[Qq]_{\bar{\bm 3}_c}[\bar Q\bar q]_{\bm 3_c}\rangle_{{\mathcal C}}+\beta |(Q\bar Q)_{\bm 1_c}(q \bar q)_{\bm 1_c}\rangle_{{\mathcal O}} + \gamma |(Q\bar q)_{\bm 1_c}(\bar Q q)_{\bm 1_c}\rangle_{{\mathcal O}} 
\label{superp}
\end{equation}
where ket subscripts ${\mathcal C},{\mathcal O}$ indicate `closed' and `open' channels respectively.
The relative size of $\alpha,\beta,\gamma$ coefficients is unknown. 
However, if we assume that $|\beta|^2,|\gamma|^2 \gg |\alpha|^2$,  diquark-antidiquark states are less likely to be formed in hadronization, but a resonance could emerge as a result of the coupling between open and closed channels, with a mechanism known in nuclear and atomic physics as the Feshbach resonance formation~\cite{leggett}. This hypothesis introduces a selection rule in the diquark-antidiquark spectrum: only the levels close enough to open channel levels are observed as physical resonances~\cite{Guerrieri:2014gfa,Papinutto:2013uya}. %
When the total energy in the open channel matches the discrete energy level in the closed channel, the two hadrons in an open channel can scatter to the intermediate state in the closed channel, which subsequently decays to give back the two particles in the open channel. 
The contribution to the scattering length due to this phenomenon is of the form 
\begin{equation}
a\sim |C|\sum_n \frac{_{{\cal C}}\langle [Qq]_{\bar{\bm 3}_c}[\bar Q\bar q]_{\bm 3_c}, n | 
H_{{\cal C}{\cal O}}|(Q\bar q)_{\bm 1_c}(\bar Q q)_{\bm 1_c}\rangle_{{\mathcal O}} }{E_{{\cal O}}-E_{n}}
\label{fesh}
\end{equation}
This sum is dominated by the term which minimizes the denominator $E_{{\cal O}}-E_{n}$. The width of the resulting resonance is naturally proportional to the square root of the detuning $\Gamma\sim \sqrt{E_{n} - E_{{\cal O}}}$ for phase space arguments. Since the $X(3872)$ is the narrowest among all $XYZ$ mesons, it must have  $\nu\approx 0$, which means the highest possible hybridization between channels given the (unknown) inter-channel interaction Hamiltonian $H_{{\cal C}{\cal O}}$. The input value to fix the discrete closed channel diquark-antidiquark spectra is  the mass of the $X(3872)$: we require that the lowest $1^{++}$ state has the mass of the $X$ and this fixes the diquark mass and the spectra as in~\cite{Maiani:2004vq,Faccini:2013lda,Maiani:2014aja}. 

The $D^+D^{*-}$ open threshold is found  to be 8~MeV heavier. Coupling between channels can give rise to a repulsive interaction if the energy of the scattering particles is larger than that of the bound state.  We might conclude that the neutral particle has no $d\bar d$ content in its wavefunction, thus explaining the well known isospin breaking pattern in  $X$ decays.  Similarly, the $X^{+}$ levels fall below $D^+\bar{D}^{*0}$ and $\bar{D}^0 D^{*+}$ thresholds, and are more difficult to be produced. 

\section{Conclusions}
To summarize, the experimental value of the prompt production cross section of the $X(3872)$ casts serious doubts on its possible interpretation in terms of a \DDstar molecule. 
The inclusion of possible interactions between comoving pions and final state mesons~\cite{Esposito:2013ada,Guerrieri:2014gfa} turned out to improve the accordance between the simulated distributions and the experimental ones and hence should be taken into account in future works.
In any case, Monte Carlo simulations leave little room for this, even when some kind of rescattering mechanism is taken into account. 
The comparison of light nuclei with $X$ production at $p_\perp$ as high as 15~GeV, suggest they do not share the same nature~\cite{Esposito:2015fsa}, but for an unbiased and definitive answer on this, 
deuteron (or hypertriton) should be searched in $pp$ collisions rather than in Pb-Pb to avoid the complications 
of subtracting medium effects. These analyses can be performed by \alice and \lhcb during Run~II.
On the other hand, the presence of open charm thresholds might enforce a Feshbach resonant mechanism~\cite{Guerrieri:2014gfa,Papinutto:2013uya}, which might explain why many states predicted by the tetraquark model are not observed in nature.

\section{ACKNOWLEDGMENTS}
I wish to thank A.~Esposito, A.~L.~Guerrieri, C.~Hanhart, L.~Maiani, F.~Piccinini, A.~D.~Polosa, V.~Riquer and \mbox{U.~Tamponi} for many precious discussions and comments. 

\bibliographystyle{aipnum-cp}%
\bibliography{tesi-nodoi}%

%Control: year  (1) truncated
\begin{thebibliography}{28}%
\makeatletter
\providecommand \@ifxundefined [1]{%
 \@ifx{#1\undefined}
}%
\providecommand \@ifnum [1]{%
 \ifnum #1\expandafter \@firstoftwo
 \else \expandafter \@secondoftwo
 \fi
}%
\providecommand \@ifx [1]{%
 \ifx #1\expandafter \@firstoftwo
 \else \expandafter \@secondoftwo
 \fi
}%
\providecommand \natexlab [1]{#1}%
\providecommand \enquote  [1]{``#1''}%
\providecommand \bibnamefont  [1]{#1}%
\providecommand \bibfnamefont [1]{#1}%
\providecommand \citenamefont [1]{#1}%
\providecommand \href@noop [0]{\@secondoftwo}%
\providecommand \href [0]{\begingroup \@sanitize@url \@href}%
\providecommand \@href[1]{\@@startlink{#1}\@@href}%
\providecommand \@@href[1]{\endgroup#1\@@endlink}%
\providecommand \@sanitize@url [0]{\catcode `\\12\catcode `\$12\catcode
  `\&12\catcode `\#12\catcode `\^12\catcode `\_12\catcode `\%12\relax}%
\providecommand \@@startlink[1]{}%
\providecommand \@@endlink[0]{}%
\providecommand \url  [0]{\begingroup\@sanitize@url \@url }%
\providecommand \@url [1]{\endgroup\@href {#1}{\urlprefix }}%
\providecommand \urlprefix  [0]{URL }%
\providecommand \Eprint [0]{\href }%
\providecommand \doibase [0]{http://dx.doi.org/}%
\providecommand \selectlanguage [0]{\@gobble}%
\providecommand \bibinfo  [0]{\@secondoftwo}%
\providecommand \bibfield  [0]{\@secondoftwo}%
\providecommand \translation [1]{[#1]}%
\providecommand \BibitemOpen [0]{}%
\providecommand \bibitemStop [0]{}%
\providecommand \bibitemNoStop [0]{.\EOS\space}%
\providecommand \EOS [0]{\spacefactor3000\relax}%
\providecommand\BibitemShut  [1]{\csname bibitem#1\endcsname}%
\let\auto@bib@innerbib\@empty
%</preamble>
\bibitem [{\citenamefont {Esposito}\ \emph
  {et~al.}(2015{\natexlab{a}})\citenamefont {Esposito}, \citenamefont
  {Guerrieri}, \citenamefont {Piccinini}, \citenamefont {Pilloni},\ and\
  \citenamefont {Polosa}}]{Esposito:2014rxa}%
  \BibitemOpen
  \bibfield  {author} {\bibinfo {author} {\bibfnamefont {A.}~\bibnamefont
  {Esposito}}, \bibinfo {author} {\bibfnamefont {A.~L.}\ \bibnamefont
  {Guerrieri}}, \bibinfo {author} {\bibfnamefont {F.}~\bibnamefont
  {Piccinini}}, \bibinfo {author} {\bibfnamefont {A.}~\bibnamefont {Pilloni}},
  \ and\ \bibinfo {author} {\bibfnamefont {A.~D.}\ \bibnamefont {Polosa}},\
  }\href@noop {} {\bibfield  {journal} {\bibinfo  {journal} {Int.J.Mod.Phys.}\
  }\textbf {\bibinfo {volume} {A30}},\ p.\ \bibinfo {pages} {1530002} (\bibinfo
  {year} {2015}{\natexlab{a}})},\ \Eprint {http://arxiv.org/abs/1411.5997}
  {arXiv:1411.5997 [hep-ph]}\BibitemShut {NoStop}%
%%CITATION = ARXIV:1411.5997;%%
\bibitem [{\citenamefont {Faccini}, \citenamefont {Pilloni},\ and\
  \citenamefont {Polosa}(2012)}]{Faccini:2012pj}%
  \BibitemOpen
  \bibfield  {author} {\bibinfo {author} {\bibfnamefont {R.}~\bibnamefont
  {Faccini}}, \bibinfo {author} {\bibfnamefont {A.}~\bibnamefont {Pilloni}}, \
  and\ \bibinfo {author} {\bibfnamefont {A.~D.}\ \bibnamefont {Polosa}},\
  }\href@noop {} {\bibfield  {journal} {\bibinfo  {journal} {Mod.Phys.Lett.}\
  }\textbf {\bibinfo {volume} {A27}},\ p.\ \bibinfo {pages} {1230025} (\bibinfo
  {year} {2012})},\ \Eprint {http://arxiv.org/abs/1209.0107} {arXiv:1209.0107
  [hep-ph]}\BibitemShut {NoStop}%
%%CITATION = ARXIV:1209.0107;%%
\bibitem [{\citenamefont {Maiani}\ \emph {et~al.}(2005)\citenamefont {Maiani},
  \citenamefont {Piccinini}, \citenamefont {Polosa},\ and\ \citenamefont
  {Riquer}}]{Maiani:2004vq}%
  \BibitemOpen
  \bibfield  {author} {\bibinfo {author} {\bibfnamefont {L.}~\bibnamefont
  {Maiani}}, \bibinfo {author} {\bibfnamefont {F.}~\bibnamefont {Piccinini}},
  \bibinfo {author} {\bibfnamefont {A.~D.}\ \bibnamefont {Polosa}}, \ and\
  \bibinfo {author} {\bibfnamefont {V.}~\bibnamefont {Riquer}},\ }\href@noop {}
  {\bibfield  {journal} {\bibinfo  {journal} {Phys.Rev.}\ }\textbf {\bibinfo
  {volume} {D71}},\ p.\ \bibinfo {pages} {014028} (\bibinfo {year} {2005})},\
  \Eprint {http://arxiv.org/abs/hep-ph/0412098} {arXiv:hep-ph/0412098 [hep-ph]}\BibitemShut {NoStop}%
%%CITATION = HEP-PH/0412098;%%
\bibitem [{\citenamefont {Maiani}\ \emph {et~al.}(2013)\citenamefont {Maiani},
  \citenamefont {Riquer}, \citenamefont {Faccini}, \citenamefont {Piccinini},
  \citenamefont {Pilloni} \emph {et~al.}}]{Faccini:2013lda}%
  \BibitemOpen
  \bibfield  {author} {\bibinfo {author} {\bibfnamefont {L.}~\bibnamefont
  {Maiani}}, \bibinfo {author} {\bibfnamefont {V.}~\bibnamefont {Riquer}},
  \bibinfo {author} {\bibfnamefont {R.}~\bibnamefont {Faccini}}, \bibinfo
  {author} {\bibfnamefont {F.}~\bibnamefont {Piccinini}}, \bibinfo {author}
  {\bibfnamefont {A.}~\bibnamefont {Pilloni}}, \ and\ \bibinfo {author}
  {\bibfnamefont {A.~D.}~\bibnamefont {Polosa}},\ }\href@noop {}
  {\bibfield  {journal} {\bibinfo  {journal} {Phys.Rev.}\ }\textbf {\bibinfo
  {volume} {D87}},\ p.\ \bibinfo {pages} {111102} (\bibinfo {year} {2013})},\
  \Eprint {http://arxiv.org/abs/1303.6857} {arXiv:1303.6857 [hep-ph]}\BibitemShut {NoStop}%
%%CITATION = ARXIV:1303.6857;%%
\bibitem [{\citenamefont {Maiani}\ \emph {et~al.}(2014)\citenamefont {Maiani},
  \citenamefont {Piccinini}, \citenamefont {Polosa},\ and\ \citenamefont
  {Riquer}}]{Maiani:2014aja}%
  \BibitemOpen
  \bibfield  {author} {\bibinfo {author} {\bibfnamefont {L.}~\bibnamefont
  {Maiani}}, \bibinfo {author} {\bibfnamefont {F.}~\bibnamefont {Piccinini}},
  \bibinfo {author} {\bibfnamefont {A.}~\bibnamefont {Polosa}}, \ and\ \bibinfo
  {author} {\bibfnamefont {V.}~\bibnamefont {Riquer}},\ }\href@noop {}
  {\bibfield  {journal} {\bibinfo  {journal} {Phys.Rev.}\ }\textbf {\bibinfo
  {volume} {D89}},\ p.\ \bibinfo {pages} {114010} (\bibinfo {year} {2014})},\
  \Eprint {http://arxiv.org/abs/1405.1551} {arXiv:1405.1551 [hep-ph]}\BibitemShut {NoStop}%
%%CITATION = ARXIV:1405.1551;%%
\bibitem [{\citenamefont {Aaij}\ \emph {et~al.}(2012)\citenamefont {Aaij} \emph
  {et~al.}}]{Aaij:2011sn}%
  \BibitemOpen
  \bibfield  {author} {\bibinfo {author} {\bibfnamefont {R.}~\bibnamefont
  {Aaij}} \emph {et~al.} (\bibinfo {collaboration} {LHCb}),\ }\href@noop {}
  {\bibfield  {journal} {\bibinfo  {journal} {Eur.Phys.J.}\ }\textbf {\bibinfo
  {volume} {C72}},\ p.\ \bibinfo {pages} {1972} (\bibinfo {year} {2012})},\
  \Eprint {http://arxiv.org/abs/1112.5310} {arXiv:1112.5310 [hep-ex]}\BibitemShut {NoStop}%
%%CITATION = ARXIV:1112.5310;%%
\bibitem [{\citenamefont {Chatrchyan}\ \emph {et~al.}(2013)\citenamefont
  {Chatrchyan} \emph {et~al.}}]{Chatrchyan:2013cld}%
  \BibitemOpen
  \bibfield  {author} {\bibinfo {author} {\bibfnamefont {S.}~\bibnamefont
  {Chatrchyan}} \emph {et~al.} (\bibinfo {collaboration} {CMS}),\ }\href@noop
  {} {\bibfield  {journal} {\bibinfo  {journal} {JHEP}\ }\textbf {\bibinfo
  {volume} {1304}},\ p.\ \bibinfo {pages} {154} (\bibinfo {year} {2013})},\
  \Eprint {http://arxiv.org/abs/1302.3968} {arXiv:1302.3968 [hep-ex]}\BibitemShut {NoStop}%
%%CITATION = ARXIV:1302.3968;%%
\bibitem [{\citenamefont {Abazov}\ \emph {et~al.}(2004)\citenamefont {Abazov}
  \emph {et~al.}}]{Abazov:2004kp}%
  \BibitemOpen
  \bibfield  {author} {\bibinfo {author} {\bibfnamefont {V.}~\bibnamefont
  {Abazov}} \emph {et~al.} (\bibinfo {collaboration} {\Dzero}),\ }\href@noop {}
  {\bibfield  {journal} {\bibinfo  {journal} {Phys.Rev.Lett.}\ }\textbf
  {\bibinfo {volume} {93}},\ p.\ \bibinfo {pages} {162002} (\bibinfo {year}
  {2004})},\ \Eprint {http://arxiv.org/abs/hep-ex/0405004}
  {arXiv:hep-ex/0405004 [hep-ex]}\BibitemShut {NoStop}%
%%CITATION = HEP-EX/0405004;%%
\bibitem [{\citenamefont {\cdf II~Collaboration}()}]{cdfnote}%
  \BibitemOpen
  \bibfield  {author} {\bibinfo {author} {\bibnamefont {\cdf
  II~Collaboration}},\ }\href@noop {} {\ }\bibinfo {note} {Note 7159, The
  ``Lifetime'' Distribution of $X(3872)$ Mesons Produced in $\bar p p$
  Collisions at \cdf}\BibitemShut {NoStop}%
\bibitem [{\citenamefont {Tomaradze}\ \emph {et~al.}(2015)\citenamefont
  {Tomaradze}, \citenamefont {Dobbs}, \citenamefont {Xiao},\ and\ \citenamefont
  {Seth}}]{Tomaradze:2015cza}%
  \BibitemOpen
  \bibfield  {author} {\bibinfo {author} {\bibfnamefont {A.}~\bibnamefont
  {Tomaradze}}, \bibinfo {author} {\bibfnamefont {S.}~\bibnamefont {Dobbs}},
  \bibinfo {author} {\bibfnamefont {T.}~\bibnamefont {Xiao}}, \ and\ \bibinfo
  {author} {\bibfnamefont {K.~K.}\ \bibnamefont {Seth}},\ }\href@noop {}
  {\bibfield  {journal} {\bibinfo  {journal} {Phys.Rev.}\ }\textbf {\bibinfo
  {volume} {D91}},\ p.\ \bibinfo {pages} {011102} (\bibinfo {year} {2015})},\
  \Eprint {http://arxiv.org/abs/1501.01658} {arXiv:1501.01658 [hep-ex]}\BibitemShut {NoStop}%
%%CITATION = ARXIV:1501.01658;%%
\bibitem [{\citenamefont {Polosa}(2015)}]{Polosa:2015tra}%
  \BibitemOpen
  \bibfield  {author} {\bibinfo {author} {\bibfnamefont {A.}~\bibnamefont
  {Polosa}},\ }\href@noop {} {\bibfield  {journal} {\bibinfo  {journal}
  {Phys.Lett.}\ }\textbf {\bibinfo {volume} {B746}},\ \unskip\ \bibinfo {pages}
  {248--250} (\bibinfo {year} {2015})},\ \Eprint
  {http://arxiv.org/abs/1505.03083} {arXiv:1505.03083 [hep-ph]}\BibitemShut
  {NoStop}%
%%CITATION = ARXIV:1505.03083;%%
\bibitem [{\citenamefont {Bignamini}\ \emph {et~al.}(2009)\citenamefont
  {Bignamini}, \citenamefont {Grinstein}, \citenamefont {Piccinini},
  \citenamefont {Polosa},\ and\ \citenamefont {Sabelli}}]{Bignamini:2009sk}%
  \BibitemOpen
  \bibfield  {author} {\bibinfo {author} {\bibfnamefont {C.}~\bibnamefont
  {Bignamini}}, \bibinfo {author} {\bibfnamefont {B.}~\bibnamefont
  {Grinstein}}, \bibinfo {author} {\bibfnamefont {F.}~\bibnamefont
  {Piccinini}}, \bibinfo {author} {\bibfnamefont {A.}~\bibnamefont {Polosa}}, \
  and\ \bibinfo {author} {\bibfnamefont {C.}~\bibnamefont {Sabelli}},\
  }\href@noop {} {\bibfield  {journal} {\bibinfo  {journal} {Phys.Rev.Lett.}\
  }\textbf {\bibinfo {volume} {103}},\ p.\ \bibinfo {pages} {162001} (\bibinfo
  {year} {2009})},\ \Eprint {http://arxiv.org/abs/0906.0882} {arXiv:0906.0882
  [hep-ph]}\BibitemShut {NoStop}%
%%CITATION = ARXIV:0906.0882;%%
\bibitem [{\citenamefont {Artoisenet}\ and\ \citenamefont
  {Braaten}(2010)}]{Artoisenet:2009wk}%
  \BibitemOpen
  \bibfield  {author} {\bibinfo {author} {\bibfnamefont {P.}~\bibnamefont
  {Artoisenet}}\ and\ \bibinfo {author} {\bibfnamefont {E.}~\bibnamefont
  {Braaten}},\ }\href@noop {} {\bibfield  {journal} {\bibinfo  {journal}
  {Phys.Rev.}\ }\textbf {\bibinfo {volume} {D81}},\ p.\ \bibinfo {pages}
  {114018} (\bibinfo {year} {2010})},\ \Eprint {http://arxiv.org/abs/0911.2016}
  {arXiv:0911.2016 [hep-ph]}\BibitemShut {NoStop}%
%%CITATION = ARXIV:0911.2016;%%
\bibitem [{\citenamefont {Bignamini}\ \emph {et~al.}(2010)\citenamefont
  {Bignamini}, \citenamefont {Grinstein}, \citenamefont {Piccinini},
  \citenamefont {Polosa}, \citenamefont {Riquer} \emph
  {et~al.}}]{Bignamini:2009fn}%
  \BibitemOpen
  \bibfield  {author} {\bibinfo {author} {\bibfnamefont {C.}~\bibnamefont
  {Bignamini}}, \bibinfo {author} {\bibfnamefont {B.}~\bibnamefont
  {Grinstein}}, \bibinfo {author} {\bibfnamefont {F.}~\bibnamefont
  {Piccinini}}, \bibinfo {author} {\bibfnamefont {A.}~\bibnamefont {Polosa}},
  \bibinfo {author} {\bibfnamefont {V.}~\bibnamefont {Riquer}}, \ and\ \bibinfo {author} {\bibfnamefont {C.}~\bibnamefont {Sabelli}},
  \ }\href@noop {} {\bibfield  {journal} {\bibinfo  {journal}
  {Phys.Lett.}\ }\textbf {\bibinfo {volume} {B684}},\ \unskip\ \bibinfo {pages}
  {228--230} (\bibinfo {year} {2010})},\ \Eprint
  {http://arxiv.org/abs/0912.5064} {arXiv:0912.5064 [hep-ph]}\BibitemShut
  {NoStop}%
%%CITATION = ARXIV:0912.5064;%%
\bibitem [{\citenamefont {Artoisenet}\ and\ \citenamefont
  {Braaten}(2011)}]{Artoisenet:2010uu}%
  \BibitemOpen
  \bibfield  {author} {\bibinfo {author} {\bibfnamefont {P.}~\bibnamefont
  {Artoisenet}}\ and\ \bibinfo {author} {\bibfnamefont {E.}~\bibnamefont
  {Braaten}},\ }\href@noop {} {\bibfield  {journal} {\bibinfo  {journal}
  {Phys.Rev.}\ }\textbf {\bibinfo {volume} {D83}},\ p.\ \bibinfo {pages}
  {014019} (\bibinfo {year} {2011})},\ \Eprint {http://arxiv.org/abs/1007.2868}
  {arXiv:1007.2868 [hep-ph]}\BibitemShut {NoStop}%
%%CITATION = ARXIV:1007.2868;%%
\bibitem [{\citenamefont {Guo}, \citenamefont {Mei{\ss}ner},\ and\
  \citenamefont {Wang}(2014)}]{Guo:2013ufa}%
  \BibitemOpen
  \bibfield  {author} {\bibinfo {author} {\bibfnamefont {F.-K.}\ \bibnamefont
  {Guo}}, \bibinfo {author} {\bibfnamefont {U.-G.}\ \bibnamefont
  {Mei{\ss}ner}}, \ and\ \bibinfo {author} {\bibfnamefont {W.}~\bibnamefont
  {Wang}},\ }\href@noop {} {\bibfield  {journal} {\bibinfo  {journal} {Commun.
  Theor. Phys.}\ }\textbf {\bibinfo {volume} {61}},\ \unskip\ \bibinfo {pages}
  {354--358} (\bibinfo {year} {2014})},\ \Eprint
  {http://arxiv.org/abs/1308.0193} {arXiv:1308.0193 [hep-ph]}\BibitemShut
  {NoStop}%
%%CITATION = ARXIV:1308.0193;%%
\bibitem [{\citenamefont {Guo}\ \emph {et~al.}(2014{\natexlab{a}})\citenamefont
  {Guo}, \citenamefont {Mei{\ss}ner}, \citenamefont {Wang},\ and\ \citenamefont
  {Yang}}]{Guo:2014sca}%
  \BibitemOpen
  \bibfield  {author} {\bibinfo {author} {\bibfnamefont {F.-K.}\ \bibnamefont
  {Guo}}, \bibinfo {author} {\bibfnamefont {U.-G.}\ \bibnamefont
  {Mei{\ss}ner}}, \bibinfo {author} {\bibfnamefont {W.}~\bibnamefont {Wang}}, \
  and\ \bibinfo {author} {\bibfnamefont {Z.}~\bibnamefont {Yang}},\ }\href@noop
  {} {\bibfield  {journal} {\bibinfo  {journal} {Eur.Phys.J.}\ }\textbf
  {\bibinfo {volume} {C74}},\ p.\ \bibinfo {pages} {3063} (\bibinfo {year}
  {2014}{\natexlab{a}})},\ \Eprint {http://arxiv.org/abs/1402.6236}
  {arXiv:1402.6236 [hep-ph]}\BibitemShut {NoStop}%
%%CITATION = ARXIV:1402.6236;%%
\bibitem [{\citenamefont {Guo}\ \emph {et~al.}(2014{\natexlab{b}})\citenamefont
  {Guo}, \citenamefont {Mei{\ss}ner}, \citenamefont {Wang},\ and\ \citenamefont
  {Yang}}]{Guo:2014ppa}%
  \BibitemOpen
  \bibfield  {author} {\bibinfo {author} {\bibfnamefont {F.-K.}\ \bibnamefont
  {Guo}}, \bibinfo {author} {\bibfnamefont {U.-G.}\ \bibnamefont
  {Mei{\ss}ner}}, \bibinfo {author} {\bibfnamefont {W.}~\bibnamefont {Wang}}, \
  and\ \bibinfo {author} {\bibfnamefont {Z.}~\bibnamefont {Yang}},\ }\href@noop
  {} {\bibfield  {journal} {\bibinfo  {journal} {JHEP}\ }\textbf {\bibinfo
  {volume} {05}},\ p.\ \bibinfo {pages} {138} (\bibinfo {year}
  {2014}{\natexlab{b}})},\ \Eprint {http://arxiv.org/abs/1403.4032}
  {arXiv:1403.4032 [hep-ph]}\BibitemShut {NoStop}%
%%CITATION = ARXIV:1403.4032;%%
\bibitem [{\citenamefont {Esposito}\ \emph {et~al.}(2013)\citenamefont
  {Esposito}, \citenamefont {Piccinini}, \citenamefont {Pilloni},\ and\
  \citenamefont {Polosa}}]{Esposito:2013ada}%
  \BibitemOpen
  \bibfield  {author} {\bibinfo {author} {\bibfnamefont {A.}~\bibnamefont
  {Esposito}}, \bibinfo {author} {\bibfnamefont {F.}~\bibnamefont {Piccinini}},
  \bibinfo {author} {\bibfnamefont {A.}~\bibnamefont {Pilloni}}, \ and\
  \bibinfo {author} {\bibfnamefont {A.}~\bibnamefont {Polosa}},\ }\href@noop {}
  {\bibfield  {journal} {\bibinfo  {journal} {J.Mod.Phys.}\ }\textbf {\bibinfo
  {volume} {4}},\ \unskip\ \bibinfo {pages} {1569--1573} (\bibinfo {year}
  {2013})},\ \Eprint {http://arxiv.org/abs/1305.0527} {arXiv:1305.0527
  [hep-ph]}\BibitemShut {NoStop}%
%%CITATION = ARXIV:1305.0527;%%
\bibitem [{\citenamefont {Guerrieri}\ \emph {et~al.}(2014)\citenamefont
  {Guerrieri}, \citenamefont {Piccinini}, \citenamefont {Pilloni},\ and\
  \citenamefont {Polosa}}]{Guerrieri:2014gfa}%
  \BibitemOpen
  \bibfield  {author} {\bibinfo {author} {\bibfnamefont {A.}~\bibnamefont
  {Guerrieri}}, \bibinfo {author} {\bibfnamefont {F.}~\bibnamefont
  {Piccinini}}, \bibinfo {author} {\bibfnamefont {A.}~\bibnamefont {Pilloni}},
  \ and\ \bibinfo {author} {\bibfnamefont {A.}~\bibnamefont {Polosa}},\
  }\href@noop {} {\bibfield  {journal} {\bibinfo  {journal} {Phys.Rev.}\
  }\textbf {\bibinfo {volume} {D90}},\ p.\ \bibinfo {pages} {034003} (\bibinfo
  {year} {2014})},\ \Eprint {http://arxiv.org/abs/1405.7929} {arXiv:1405.7929
  [hep-ph]}\BibitemShut {NoStop}%
%%CITATION = ARXIV:1405.7929;%%
\bibitem [{\citenamefont {Adam}\ \emph
  {et~al.}(2015{\natexlab{a}})\citenamefont {Adam} \emph
  {et~al.}}]{Adam:2015yta}%
  \BibitemOpen
  \bibfield  {author} {\bibinfo {author} {\bibfnamefont {J.}~\bibnamefont
  {Adam}} \emph {et~al.} (\bibinfo {collaboration} {ALICE}),\ }\href@noop {} {\
   (\bibinfo {year} {2015}{\natexlab{a}})},\ \Eprint
  {http://arxiv.org/abs/1506.08453} {arXiv:1506.08453 [nucl-ex]}\BibitemShut
  {NoStop}%
%%CITATION = ARXIV:1506.08453;%%
\bibitem [{\citenamefont {Adam}\ \emph
  {et~al.}(2015{\natexlab{b}})\citenamefont {Adam} \emph
  {et~al.}}]{Adam:2015vda}%
  \BibitemOpen
  \bibfield  {author} {\bibinfo {author} {\bibfnamefont {J.}~\bibnamefont
  {Adam}} \emph {et~al.} (\bibinfo {collaboration} {ALICE}),\ }\href@noop {} {\
   (\bibinfo {year} {2015}{\natexlab{b}})},\ \Eprint
  {http://arxiv.org/abs/1506.08951} {arXiv:1506.08951 [nucl-ex]}\BibitemShut
  {NoStop}%
%%CITATION = ARXIV:1506.08951;%%
\bibitem [{\citenamefont {Esposito}\ \emph
  {et~al.}(2015{\natexlab{b}})\citenamefont {Esposito}, \citenamefont
  {Guerrieri}, \citenamefont {Maiani}, \citenamefont {Piccinini}, \citenamefont
  {Pilloni}, \citenamefont {Polosa},\ and\ \citenamefont
  {Riquer}}]{Esposito:2015fsa}%
  \BibitemOpen
  \bibfield  {author} {\bibinfo {author} {\bibfnamefont {A.}~\bibnamefont
  {Esposito}}, \bibinfo {author} {\bibfnamefont {A.~L.}\ \bibnamefont
  {Guerrieri}}, \bibinfo {author} {\bibfnamefont {L.}~\bibnamefont {Maiani}},
  \bibinfo {author} {\bibfnamefont {F.}~\bibnamefont {Piccinini}}, \bibinfo
  {author} {\bibfnamefont {A.}~\bibnamefont {Pilloni}}, \bibinfo {author}
  {\bibfnamefont {A.~D.}\ \bibnamefont {Polosa}}, \ and\ \bibinfo {author}
  {\bibfnamefont {V.}~\bibnamefont {Riquer}},\ }\href@noop {} {\bibfield
  {journal} {\bibinfo  {journal} {Phys.Rev.}\ }\textbf {\bibinfo {volume}
  {D92}},\ p.\ \bibinfo {pages} {034028} (\bibinfo {year}
  {2015}{\natexlab{b}})},\ \Eprint {http://arxiv.org/abs/1508.00295}
  {arXiv:1508.00295 [hep-ph]}\BibitemShut {NoStop}%
%%CITATION = ARXIV:1508.00295;%%
\bibitem [{\citenamefont {Schnedermann}, \citenamefont {Sollfrank},\ and\
  \citenamefont {Heinz}(1993)}]{Schnedermann:1993ws}%
  \BibitemOpen
  \bibfield  {author} {\bibinfo {author} {\bibfnamefont {E.}~\bibnamefont
  {Schnedermann}}, \bibinfo {author} {\bibfnamefont {J.}~\bibnamefont
  {Sollfrank}}, \ and\ \bibinfo {author} {\bibfnamefont {U.~W.}\ \bibnamefont
  {Heinz}},\ }\href@noop {} {\bibfield  {journal} {\bibinfo  {journal}
  {Phys.Rev.}\ }\textbf {\bibinfo {volume} {C48}},\ \unskip\ \bibinfo {pages}
  {2462--2475} (\bibinfo {year} {1993})},\ \Eprint
  {http://arxiv.org/abs/nucl-th/9307020} {arXiv:nucl-th/9307020 [nucl-th]}\BibitemShut {NoStop}%
%%CITATION = NUCL-TH/9307020;%%
\bibitem [{\citenamefont {Cho}\ \emph {et~al.}(2011)\citenamefont {Cho} \emph
  {et~al.}}]{Cho:2010db}%
  \BibitemOpen
  \bibfield  {author} {\bibinfo {author} {\bibfnamefont {S.}~\bibnamefont
  {Cho}} \emph {et~al.} (\bibinfo {collaboration} {ExHIC}),\ }\href@noop {}
  {\bibfield  {journal} {\bibinfo  {journal} {Phys.Rev.Lett.}\ }\textbf
  {\bibinfo {volume} {106}},\ p.\ \bibinfo {pages} {212001} (\bibinfo {year}
  {2011})},\ \Eprint {http://arxiv.org/abs/1011.0852} {arXiv:1011.0852
  [nucl-th]}\BibitemShut {NoStop}%
%%CITATION = ARXIV:1011.0852;%%
\bibitem [{\citenamefont {Gyulassy}\ and\ \citenamefont
  {Plumer}(1990)}]{Gyulassy:1990ye}%
  \BibitemOpen
  \bibfield  {author} {\bibinfo {author} {\bibfnamefont {M.}~\bibnamefont
  {Gyulassy}}\ and\ \bibinfo {author} {\bibfnamefont {M.}~\bibnamefont
  {Plumer}},\ }\href@noop {} {\bibfield  {journal} {\bibinfo  {journal}
  {Phys.Lett.}\ }\textbf {\bibinfo {volume} {B243}},\ \unskip\ \bibinfo {pages}
  {432--438} (\bibinfo {year} {1990})}\BibitemShut {NoStop}%
%%CITATION = PHLTA,B243,432;%%
\bibitem [{\citenamefont {Leggett}(2006)}]{leggett}%
  \BibitemOpen
  \bibfield  {author} {\bibinfo {author} {\bibfnamefont {A.~J.}\ \bibnamefont
  {Leggett}},\ }\href@noop {} {\emph {\bibinfo {title} {Quantum liquids: Bose
  condensation and Cooper pairing in condensed-matter systems}}}\ (\bibinfo
  {publisher} {Oxford University Press},\ \bibinfo {year} {2006})\BibitemShut
  {NoStop}%
\bibitem [{\citenamefont {Papinutto}\ \emph {et~al.}()\citenamefont
  {Papinutto}, \citenamefont {Piccinini}, \citenamefont {Pilloni},
  \citenamefont {Polosa},\ and\ \citenamefont {Tantalo}}]{Papinutto:2013uya}%
  \BibitemOpen
  \bibfield  {author} {\bibinfo {author} {\bibfnamefont {M.}~\bibnamefont
  {Papinutto}}, \bibinfo {author} {\bibfnamefont {F.}~\bibnamefont
  {Piccinini}}, \bibinfo {author} {\bibfnamefont {A.}~\bibnamefont {Pilloni}},
  \bibinfo {author} {\bibfnamefont {A.~D.}\ \bibnamefont {Polosa}}, \ and\
  \bibinfo {author} {\bibfnamefont {N.}~\bibnamefont {Tantalo}},\ }\href@noop
  {} {\ }\Eprint {http://arxiv.org/abs/1311.7374} {arXiv:1311.7374 [hep-ph]}\BibitemShut {NoStop}%
%%CITATION = ARXIV:1311.7374;%%
\end{thebibliography}%

\end{document}